\documentclass[twoside]{article}
\usepackage{qic,amsbsy}

\newcommand\myu{u}

\begin{document}
\setlength{\textheight}{8.0truein}    

\runninghead{THE REGIME OF GOOD CONTROL}
            {J. Li and K. Jacobs}

\normalsize\textlineskip
\thispagestyle{empty}
\setcounter{page}{1}

\copyrightheading{0}{0}{2003}{000--000}

\vspace*{0.88truein}

\alphfootnote

\fpage{1}

\centerline{\bf THE EQUATIONS OF QUANTUM FEEDBACK CONTROL }
\vspace*{0.035truein}
\centerline{\bf IN THE REGIME OF GOOD CONTROL}
\vspace*{0.37truein}
\centerline{\footnotesize Juliang Li and Kurt Jacobs}
\vspace*{0.015truein}
\centerline{\footnotesize\it Department of Physics, University of Massachusetts at Boston, 100 Morrissey Blvd}
\baselineskip=10pt
\centerline{\footnotesize\it Boston, MA 02125, USA}
\vspace*{0.225truein}
\publisher{(received date)}{(revised date)}

\vspace*{0.21truein}

\abstracts{
We derive the equations of motion describing the feedback control of quantum systems in the regime of ``good control", in which the control is sufficient to keep the system close to the desired state. One can view this regime as the quantum equivalent of the ``linearized" regime for feedback control of classical nonlinear systems. Strikingly, while the dynamics of a single qubit in this regime is indeed linear, that of all larger systems remains nonlinear, in contrast to the classical case. As a first application of these equations, we determine the steady-state performance of feedback protocols for a single qubit that use unbiased measurements. 
}{}{}

\vspace*{10pt}

\vspace*{3pt}
\communicate{***}

\vspace*{1pt}\textlineskip    
\section{Introduction}
The process of real-time feedback~\cite{OLRJacobs, Whittle96, JacobsShabani09} from a continuous measurement~\cite{JacobsSteck06,Brun02} is a potentially important tool for obtaining precise control of noisy quantum systems~\cite{Smith02, Armen02, Steck04, James04, Ralph04, Combes06, Yanagisawa06, Geremia06, Matsukevich06, Bushev06, Ticozzi06, Tian07, Katz07, Cook07, Jacobs07c, Altafini07, Wang07, Sagawa08, Combes08}. In general the dynamics of a continuously observed quantum system is nonlinear~\cite{JacobsSteck06}. Because of this, it is difficult, if not impossible, to obtain analytically, or even numerically, fully optimal feedback protocols for most quantum systems. One therefore wishes to find ways to simplify the problem, yet still obtain useful results. 

In considering feedback control of nonlinear classical systems, if one assumes that the deviations from the ``target" state about which one is stabilizing the system are small, then the dynamics about this state are approximately linear. Thus the assumption of small deviations about the target state (that is, that the controller is able to effect good control) effectively ``linearizes" the dynamics, allowing the application of the optimal control results for linear systems~\cite{Maybeck3}. 

We now ask the question, if we consider an equivalent regime for quantum systems, that of ``good control", does this simplify the dynamics of the feedback control process, and might it allow us to obtain optimal feedback protocols? Two previous works have examined this regime, but only in combination with ``strong feedback" (\cite{Jacobs07c, Shabani08}, see also \cite{JacobsShabani09}). This is the additional assumption that the Hamiltonian at the disposal of the controller can induce dynamics sufficiently fast compared to that of the noise and the measurement, that it can be assumed infinite. An optimal control protocol was obtained for a qutrit in this regime in~\cite{Shabani08}. 

Since the regime of good control is one in which the majority of control systems will wish to operate, one would ideally like to drop the restriction of strong feedback, and obtain results that are applicable to all systems in the regime of good control. We achieve this by deriving the equations of motion for both the eigenvectors and eigenvalues of the density matrix under a continuous measurement of an arbitrary observable. The regime of good control can then be specified in terms of these eigenvalues and eigenvectors. 

As a first application of these equations we consider the simplest case, that of a single qubit, and derive an expression for the steady-state performance of feedback when the measurement is continually adapted so that the measured observable remains unbiased with respect to the density matrix. This has been shown to be the optimal measurement strategy in the limit of strong feedback~\cite{rapidP, Wiseman06x, Wiseman07, Shabani08}. In the steady-state, the performance of the feedback algorithm is quantified by the steady-state probability, $P_{\mbox{\scriptsize ss}}$, that the system will be found in the desired state (the target state). We derive an explicit expression for $P_{\mbox{\scriptsize ss}}$ for the optimal linear feedback, for spontaneous decay and dephasing of the qubit.   

In the following section we derive the equations of motion for the eigenvalues and eigenvectors of the density matrix under a continuous measurement of an observable, and use these to determine the equations of motion for feedback control in the regime of good control. In Section III we apply these to a specific problem for a single qubit. Section IV concludes with a discussion of the implications of these results.  
 
\section{The Equations of Feedback Control} 

\subsection{Eigenvectors and Eigenvalues} 

We first derive the equations of motion for the eigenvalues and eigenvectors of the density matrix under a continuous measurement of an arbitrary observable $X$. The evolution of the density matrix under such a measurement is given by the stochastic master equation (SME)~\cite{Brun02,JacobsSteck06} 
\begin{eqnarray}
  d\rho & = &  - k[X,[X, \rho ]] dt + \sqrt{2k}( X\rho+ \rho X - 2\langle X \rangle \rho) dW   ,
           \label{eq1}
\end{eqnarray} 
where $\rho$ is the system density matrix, and $k$, often referred to as the {\em measurement strength}, characterizes the rate at which the measurement extracts information. 

To obtain the equations of motion for the eigenvalues and eigenvectors, we need to know how they change to first order in $dt$. It turns out that time-independent perturbation theory is the perfect tool for this task. The density matrix at time $t+dt$ is 
\begin{eqnarray}
  \rho(t+dt) = \rho(t) + d\rho . 
           \label{eq1}
\end{eqnarray} 
Recall that time-independent perturbation theory is a method for determining the eigenvalues and eigenvectors of a Hamiltonian $H$, when  
\begin{eqnarray}
   H = H_0 + \lambda V , 
\end{eqnarray}
the eigenvalues and eigenvectors of $H_0$ are known, and $\lambda$ is a sufficiently small parameter~\cite{Bransden}. The solution is given as a power series in $\lambda$.
   
Identifying $\rho(t)$ with $H_0$ and $d\rho$ with $\lambda V$, time-independent perturbation theory gives us the eigenvalues and eigenvectors of $\rho(t+dt)$ in terms of $\rho(t)$, which is what we need. In calculating these eigenvectors and eigenvalues, it is important to note that since $d\rho$ includes the stochastic It\^{o} increment, $dW$, and since $dW^2 = dt$, we must use the perturbation theory to {\em second order} in $\lambda$. Note that while second-order perturbation theory would usually give an approximation to the true dynamics, in this case it generates the exact equations of motion for the eigenvectors and eigenvalues of $\rho$. 

In the following we will denote the eigenvalues of $\rho(t)$ as $\lambda_n (t)$. We will also write the eigenvectors of $\rho$ at time $t+dt$ in terms of the eigenvectors of $\rho$ at time $t$. Denoting the former by $|n(t+dt)\rangle$, and the latter as $|n(t)\rangle$, we thus write 
\begin{equation}
|n(t+dt)\rangle = \sum_{j=0}^{N-1} ( \delta_{nj} + dc_{jn} ) |j(t)\rangle . 
\end{equation}
Finally, we will denote the matrix elements of $X$ in the eigenbasis of $\rho(t)$ as $X_{jk}$; that is, $X_{jk} \equiv \langle j(t) | X | k(t) \rangle$. With this notation the equations of motion are 
\begin{eqnarray}
    d \lambda_n & = & 8k dt  \left[ \sum_{l\not=n}  \frac{ \lambda_l }{\lambda_n - \lambda_l} |X_{nl}|^2 \right] \lambda_n   +  \sqrt{8k} dW\left( X_{nn} - \sum_{l} \lambda_l X_{ll}  \right)  \lambda_n  
    \label{EVeqs1} \\ 
    d c_{jn} & = &  4 k dt \left[  X_{jn}  \left( \left\{ \frac{1}{4} - \frac{\lambda_n (\lambda_n + \lambda_j)}{(\lambda_n - \lambda_j)^2} \right\} X_{nn}  + \frac{\lambda_n^2 - \lambda_j^2}{(\lambda_n - \lambda_j)^2} \sum_{l}\lambda_l X_{ll} \right)  \right]  \nonumber  \\  
    & & - k dt \left[  \sum_{l\not=n} \left( \frac{\lambda_j + \lambda_n - 2\lambda_l}{(\lambda_n - \lambda_j)}  - \frac{2 (\lambda_j + \lambda_l) (\lambda_l + \lambda_n)}{(\lambda_n - \lambda_l)(\lambda_n - \lambda_j)}   \right) X_{jl}X_{ln} \right]  \nonumber \\
     & &  + \sqrt{2k} dW \left( \frac{\lambda_n + \lambda_j}{\lambda_n - \lambda_j} \right) X_{jn}             , \;\;\; n \not= j \label{EVeqs2} \\
    d c_{nn} & = & - k dt \left[ \sum_{l\not=n} \frac{(\lambda_n + \lambda_l)^2}{(\lambda_n - \lambda_l)^2} |X_{nl}|^2 \right] \label{EVeqs3} . 
\end{eqnarray}
This set of equations makes explicit the complexity of the dynamics of continuous measurement. 

\subsection{Feedback Control} 

We now turn to feedback control, in which the controller continually modifies the Hamiltonian of the system, based upon the measurement results, in order to bring the state of the system close to a ``target" state $|\psi\rangle$. The target state could change with time, but we will restrict ourselves here to a fixed target state. This loses little, as it is straightforward to modify the following equations to take into account an evolving target state. We will denote the total Hamiltonian of the system by $H$. Usually $H$ breaks down naturally into the sum $H_0 + H_{\mbox{\scriptsize fb}}(t)$, where $H_{\mbox{\scriptsize fb}}(t)$ is the part that the controller can modify. At time $t$, the Hamiltonian $H_{\mbox{\scriptsize fb}}(t)$, and thus $H(t)$, is some function of the measurement record up until that time. It can be shown that optimal control can always be realized by choosing $H(t)$ to be a function of $\rho(t)$ (which is itself obtained from the measurement record)~\cite{DHJMT}.  

While it is the target state that remains fixed, and the system density matrix that evolves, one can always view the dynamics as happening the other way around, since the first case is related to the second by a unitary transformation. With this picture in mind, we will write the target state, $|\psi\rangle$, in terms of the eigenvectors of $\rho(t)$. Thus 
\begin{equation} 
   |\psi\rangle = \sum_{k=0}^{N-1} z_n(t) |n(t)\rangle . 
   \label{target}
\end{equation} 
The equations of motion above for the $c_{jn}$ allow us to determine the equations of motion of the $z_n$ under the continuous measurement. To do so we note that since the transformation from $|n(t)\rangle$ to $|n(t+dt)\rangle$ is unitary (to first order in $dt$), the inverse transformation is given by the Hermitian conjugate: $|n(t)\rangle = \sum_{j=0}^{N-1} ( \delta_{nj} + dc_{nj}^* ) |j(t+dt)\rangle$. Substituting this into Eq.(\ref{target}) gives the equations of motion for the $z_n$, which are 
\begin{equation}
   dz_n = \sum_j  dc_{jn}^* z_j . 
\end{equation}

To describe feedback control we must include, in addition to the dynamics induced by the measurement, the dynamics induced by the noise from the environment, as this is the reason that one needs to use feedback control in the first place, and the dynamics due to the Hamiltonian, $H(t)$. The noise effects both the eigenvalues of the density matrix as well as the $z_n$, but the Hamiltonian, and thus the feedback, only influences the $z_n$.

Environmental noise is usually described well by the Lindblad master equation~\cite{Wiseman01} 
\begin{equation}
  \dot{\rho}_{\mbox{\scriptsize env}} = -\gamma \left( L^\dagger L \rho + \rho L^\dagger L - 2 L \rho L^\dagger  \right) ,  
  \label{lindblad1}
\end{equation}
where $L$ is an arbitrary operator. We can have as many terms of the above form as we wish, so that the effect of the environment is more generally given by  
\begin{equation}
  \dot{\rho}_{\mbox{\scriptsize env}} = - \sum_{m} \gamma_m \left( L^{(m)\dagger} L^{(m)} \rho + \rho L^{(m)\dagger} L^{(m)} - 2 L^{(m)} \rho L^{(m)\dagger} \right) . 
\end{equation}
Using time-independent perturbation theory as above, the dynamics induced by Eq.(\ref{lindblad1}) is 
\begin{eqnarray}
    \dot{\lambda}_n & = & -2 \gamma \left[ \lambda_n D_{nn}   
    				- \sum_k \lambda_k |L_{nk}|^2 \right] , \label{deco1} \\ 
     \dot{c}_{jn} & = & - \gamma \left[ \frac{(\lambda_n + \lambda_j) D_{jn}}{\lambda_n -\lambda_j }   - 2 \sum_k  \frac{\lambda_k L_{jk}^* L_{nk}}{\lambda_n - \lambda_j}  \right] , \label{deco2}  \\  
    \dot{c}_{nn} & = & 0 ,  \label{deco3}
\end{eqnarray}
where we have defined $D_{jn} = \sum_{k} L_{kj}^* L_{kn}$. Finally, the dynamics induced by the Hamiltonian is  
\begin{equation} 
   \dot{c}_{jn} =   i H_{jn} . 
\end{equation} 

The quantity that we want feedback control to maximize is the probability, $P$, that the system will be found in the target state. This is 
\begin{equation}
   P = \langle \psi | \rho(t) |\psi\rangle = \sum_{n} |z_n|^2 \lambda_n. 
\end{equation}
The performance of feedback control is thus completely described by the evolution of the $\lambda_n$ and the $z_n$. However, this evolution may also depend upon the dynamics of the elements of the eigenvalues not captured by the $z_n$. This is because the equations of motion for the $\lambda_n$ and $z_n$ depend on the noise operator $L$, and all our operators are written in the basis of the density matrix. As this basis evolves, the matrix elements of $L$ also undergo an effective evolution. (The matrix elements of $H$ and $X$ will also evolve in the same way, if they are not being continually specified by the controller.) The equation of motion for the elements of an arbitrary operator $A$ in the density matrix eigenbasis are 
\begin{equation}
    dL_{mn} =  \sum_j (L_{jn} dc_{mj} + L_{mj} dc_{nj}^* ) +  \sum_{jk} L_{jk} dc_{mj} dc_{nk}^* .   \label{eq:gcfin}
\end{equation}
So, in addition to the motion of $\lambda_n$ and $z_n$, we must include the evolution of the elements of the noise operator(s) in our feedback control problem. In general this also means that the $\lambda_n$ are coupled to the $z_n$. In the most general case, including the effective motion of the decoherence operators is equivalent to including the motion of all the eigenvectors of the density matrix. It is nevertheless useful to write the operators, as we have above, in the density matrix eigenbasis. This is because in certain cases, for example if the noise is isotropic (the same in all directions in Hilbert space), the noise is basis independent, and so the problem reduces entirely to the dynamics of the $\lambda_n$ and $z_n$.

\subsection{Specifying the Regime of Good Control} 

The regime of good control is defined as that in which the feedback protocol keeps the system very close to the target state throughout the evolution. That is, the probability that the system will be found in the target state, $P$, is close to unity throughout the evolution. This is a regime that many feedback control systems will wish to operate in, especially if they are designed to operate in the steady state, and is thus an important regime. The utility of writing the equations of motion of feedback control in terms of the eigenvectors and eigenvalues of $\rho$ is that it allows us make the approximation $P = 1 - \varepsilon$, with $\varepsilon \ll 1$. To do this we first note that $P$ can only be close to unity if both $z_n$ and $P_n$ are close to unity, for the same value of the index $n$. Ordering the eigenvalues in decreasing order, this means that $\lambda_0 = 1 - \Delta$, $z_0 = 1 - \delta$, where $\Delta$ and $\delta$ are small.  (Note that here we have assumed that $z_0$ is real. We will justify this assumption below.)  

We now choose to specify the regime of good control so that the small eigenvalues, and the small coefficients $z_n$, $n\geq 1$, are of the same order. Since $|z_0|^2 + \sum_{n=1}^{N} |z_n|^2 = 1$, this means that, to first order in $\delta$ 
\begin{equation}
   \delta =   \frac{1}{2} |z_n|^2 \sim \Delta^2 ,  \;\;\;\;\;\;  n \geq 1 . 
\end{equation}
This means that, in expanding the equations of motion to first order in $\Delta$, we can set $\delta = 0$.  This is not the only choice we can make, but it is the one that results in equations of motion that are linear in the $z_n$. We will return to discuss this choice in the final section of this paper. The probability that the system is in the target state, $P$, becomes    
\begin{eqnarray}
   P  \approx |z_0|^2 \lambda_0 \approx  1 - \Delta . 
\end{eqnarray}
Now we see why we can assume that $z_0$ is real: if $\delta=0$, then $z_0$ is constant, and we are free to choose the global phase of the target state so that $z_0$ is initially real.  

We now expand the equations of motion of the eigenvalues to first order in $\Delta$, and the result is  
\begin{eqnarray} 
   d\lambda_{n} & = &  - \lambda_n  \left\{ 8 k dt \left[  |X_{0n}|^2 +
\sum_{j>0, j\not= n}\frac{\lambda_j}{\lambda_n - \lambda_j}
|X_{jn}|^2 \right] +  \sqrt{8k} dW ( X_{00} - X_{nn}) \right\}   
      \label{eq:gc1} 
\end{eqnarray} 
for $n \geq 1$, and 
\begin{eqnarray}
  d\Delta & = & \sum_{n\not=0} d\lambda_n  = - 8 k dt \sum_{n\not=0} \lambda_n  |X_{0n}|^2 + \sqrt{8k} dW \left( \Delta X_{00} - \sum_{n\not=0} \lambda_n X_{nn} \right)  .
\end{eqnarray} 
The purpose of expanding the equations of motion to first order in $\Delta$ is to simplify them. Since the equations of motion for the $z_n$ are already linear in the $z_n$, the wisdom of applying the expansion to them is not so clear. However, this expansion does do two things. The first is to reduce the number of variables by eliminating $z_0$. The second is that it removes some nonlinear terms of the form $\lambda_j z_k$. The resulting equations of motion for the $z_n$ are     
\begin{eqnarray}
   dz_{n} & = &  k dt \left[ X_{00} X_{0n} - \sum_{j\not=0} X_{0j} X_{jn}\right] 
   			+ \sqrt{2k} dW X_{0n} \nonumber \\ 
	        & & - \lambda_n \left\{    k dt  \left[  4 X_{00} + \sum_{j\not=0} X_{0j} X_{jn} \right]   - \sqrt{8k} dW X_{0n}   \right\}  \nonumber \\ 
		& &   - \sum_{j\not=0} \lambda_j \left\{    k dt  \left[  4 X_{0n}(X_{00} - X_{jj}) + X_{0j} X_{jn} - 2 \sum_{k\not=0} X_{0k} X_{kn}  \right]  + \sqrt{8k} dW X_{0n}  \right\}  \nonumber \\
		& & 	  - z_n \left\{  k dt \left[  |X_{0n}|^2  +  \sum_{j\not=0,n} \frac{(\lambda_n + \lambda_j)^2}{(\lambda_n - \lambda_j)^2} |X_{nj}|^2 \right] \right\} + \sum_{j\not=0,n} z_j F_{nj}( \boldsymbol{\lambda} , X ) 
\end{eqnarray} 
for $n \geq 1$, where  
\begin{eqnarray} 
  F_{nj}( \boldsymbol{\lambda} , X ) & = &  - k dt  \left[  \frac{3(\lambda_j + \lambda_n)}{(\lambda_j - \lambda_n)} \right]  X_{j0}X_{0n}  \nonumber \\
       & & - k dt \sum_{k\not=0,j} \left[ \frac{(\lambda_n - \lambda_k) + (\lambda_j - \lambda_k)}{(\lambda_j - \lambda_n)} - \frac{2(\lambda_n + \lambda_k)(\lambda_j + \lambda_k)}{(\lambda_j - \lambda_n)(\lambda_j - \lambda_k)}\right]  X_{j0}X_{0n} \nonumber \\ 
  & & - 4 k dt \left[  \left( \frac{\lambda_j(\lambda_j + \lambda_n) }{(\lambda_j - \lambda_n)^2} - \frac{1}{4} \right)  X_{jj}X_{nj} - \frac{(\lambda_j^2 - \lambda_n^2)}{(\lambda_j - \lambda_n)^2} X_{00}X_{jn} \right] \nonumber \\ 
  & & + \sqrt{2k} dW \left( \frac{\lambda_j + \lambda_n}{\lambda_j - \lambda_n} \right) X_{jn} .
\end{eqnarray}

In the regime of good control, the dynamics due to the environment (Eqs.(\ref{deco1}) - (\ref{deco3})) becomes  
\begin{eqnarray} 
 \dot{\lambda}_{n} \!\! & = &  \!\! 2 \gamma \left[ |L_{n0}|^2 -  \lambda_n  D_{nn} -  \sum_{j\not= 0} \lambda_j ( |L_{n0}|^2 - |L_{nj}|^2) \right]  , \\
  \dot{z}_n \!\! & = & \!\! -\gamma \left[ D_{n0}  - 2 L_{00}^* L_{n0}    + \sum_{j\not= 0} z_j  D_{nj}  + 2   \lambda_n  D_{n0}  - 2 \sum_{j\not= 0} \lambda_j ( L_{0j}^* L_{nj} - L_{00}^* L_{n0} ) \right] , \;\;
\end{eqnarray}
where, as above, $D_{jn} = \sum_{k} L_{kj}^* L_{kn}$. Finally, the dynamics due to the Hamiltonian is  
\begin{eqnarray}
   \dot{z}_n & = & - iH_{n0} - i\sum_{j\not=0} H_{nj} z_j  . 
\end{eqnarray}

When using the above equations, Eqs.(\ref{eq:gc1})-(\ref{eq:gcfin}), it is important to remember that the elements of all operators, the measured observable, $X$, the Hamiltonian, $H$, and the decoherence (Lindblad) operator $L$, are given in the basis of the eigenvectors of $\rho$. So long as $X$ and $H$ are determined by the feedback protocol, it does not matter which basis one writes them in. However, for the noise operator $L$, and for the measured observable if it is fixed, then the elements of these operators are not constant in the density matrix eigenbasis, but are determined by the evolution of the eigenvectors, as describe above. 

\section{A Single Qubit: Feedback Control with Unbiased Measurements} 

To calculate the steady-state performance of a feedback protocol, we need to solve for the steady-state values of $\lambda_n$, averaged over all trajectories. The differential equations for $\lambda_n$ are in general coupled to those for the $z_n$. While the equations of motion for the $z_n$ are linear, those for the eigenvalues are only linear if the system has two states. In addition, if the dimension of the system is larger than two, in general the equations of motion have multiplicative noise (the noise multiplies the variables themselves). For linear systems driven by additive noise, general and exact results exist for feedback control, providing both optimal~\cite{Whittle96, Maybeck3} and robust~\cite{ZhouDoyle97} control protocols. So we see that in the regime of good control, these exact results can be applied to a single qubit, but not to higher dimensional systems.   

We will now use the equations derived in the previous section to calculate the performance of feedback control for a single qubit, where the measured observable is continually adjusted so that it is always unbiased with respect to the eigenbasis of the density matrix. This kind of ``unbiased" measurement has a special property; it eliminates the stochastic terms in the equations of motion for the eigenvalues. To see why we first note that the diagonal matrix elements of an observable that is unbiased w.r.t the density matrix eigenbasis are all equal in this basis. Further, the master equation that describes the measurement of $X$, Eq.(\ref{eq1}), is invariant under the transformation $X \rightarrow X + \alpha I$, where $I$ is the identity operator and $\alpha$ is any real number. This means that we can always choose $X$ so that $\mbox{Tr}[X] = 0$. With this choice, the diagonal elements of an unbiased $X$ are all zero, and this   eliminates the stochastic terms from equations for $\lambda_n$.  An unbiased measurement has also been shown to be the optimal measurement for feedback control in the regime of strong feedback~\cite{Shabani08}. 

For a single qubit, since the diagonal elements of $X$ are zero, $X$ is specified by only one complex number, $X_{01}$. An examination of the eqation of motion for $z_1$ shows that the phase of this complex number only serves to redistribute the noise between the real and imaginary parts of $z_1$, and thus does not affect the resulting performance of the feedback protocol, we chose $X_{01}$ to be real. This means that we can completely absorb $X_{01}$ into the measurement strength $k$, and thus set $X_{01} = 1$. The equations of motion for the qubit in the regime of good control are then 
\begin{eqnarray}
  \!\!\!\! \!\!\!\! \!\!\!\! \dot{\lambda}_1 & = & 2 \gamma L_{10}^2  - (8 k + 2 \gamma [L_{01}^2 + L_{10}^2])\lambda_1  + 4 \gamma  L_{10}( L_{11} -L_{00})z_1  \label{lam1} \\ 
  \!\!\!\! \!\!\!\! \!\!\!\! dz_1 & = & \left[ - \gamma (D_{10}-2L_{00}^*L_{10})  - i \myu H_{10} \right] dt  -  2\gamma (  D_{10} - L_{01}^*L_{11} + L_{00}^*L_{10})   \lambda_1 dt \nonumber \\ 
    & &   -  ( k  + \gamma D_{11}  + i \myu H_{11} ) z_{1} dt + \sqrt{2k} dW,  \label{z1}
\end{eqnarray} 
Here we have also extracted an overall rate constant, $\myu$, from the Hamiltonian, so that the Hamiltonian matrix elements, $H_{jk}$, are now dimensionless. 

Since the density matrix  eigenbasis is determined entirely by  $z_1$, the elements of the noise operator $L$ that appear in the equations above are, to first order in $\Delta$,  
\begin{eqnarray*} 
  L_{00} & = &  \tilde{L}_{00}  + \tilde{L}_{01} z_1 + \tilde{L}_{10} z_1^* ,  \\ 
  L_{01} & = &  - \tilde{L}_{00} z_1 +  \tilde{L}_{11} z_1^*  + \tilde{L}_{01} ,  \\ 
  L_{10} & = &  -  \tilde{L}_{00} z_1^* + \tilde{L}_{11} z_1  + \tilde{L}_{10} ,  \\ 
  L_{11} & = &   \tilde{L}_{11} - \tilde{L}_{01} z_1^* - \tilde{L}_{10} z_1 ,  
\end{eqnarray*} 
where the $\tilde{L}_{jk}$ are constant. 

We wish to obtain the steady-state solutions to Eqs.(\ref{lam1}) and (\ref{z1}). We now simplify the analysis by specializing to the important and widely applicable case in which  the qubit undergoes both spontaneous decay at rate $\Gamma$, for which   
\begin{equation}
    \tilde{L} = \left(  \begin{array}{cc}
		  0  &  0   \\
		  1  &  0
		  \end{array}  \right) , 
\end{equation}
and dephasing at rate $\gamma$, for which   
\begin{equation}
    \tilde{L} = \left(  \begin{array}{cc}
		  0  &  1   \\
		  1  &  0 
		  \end{array}  \right) .  
\end{equation}
The dynamical equations for the feedback control problem reduce to  
\begin{eqnarray}
  \dot{\lambda}_1 & = & - 2 (4 k +  \Gamma + 2 \gamma) \lambda_1 + 2 (\Gamma + \gamma), \\  
  dz_1 & = & -(k + \gamma - i \myu H_{11} ) z_{1} dt - i \myu H_{10} dt + \sqrt{2k} dW , 
  \label{z1de}
\end{eqnarray}
Now the equations have simplified, in that $\lambda_1$ is decoupled from $z_1$. We can immediately obtain the steady-state value of $\lambda_1$, which is 
\begin{equation}
\lambda_1^{\mbox{\scriptsize ss}} = \frac{\Gamma + \gamma}{4 k + \Gamma + 2 \gamma} .   
   \label{lamss}
\end{equation} 
The performance, in this case of {\em all} feedback control algorithms that use unbiased measurements, in the regime of good control, is therefore   
\begin{equation} 
   P_{\mbox{\scriptsize ss}} =  1- \lambda_1^{\mbox{\scriptsize ss}} =  \frac{1}{1 + \frac{ \Gamma + \gamma}{4k + \gamma}}.    
\end{equation}  
However, this performance will only be achieved so long as the feedback is sufficient to keep the system in the regime of good control. To determine the conditions under which this is true we must solve the equation of motion for $z_1$. Since this equation is stochastic, this means determining the steady-state values of the mean and variance of $z_1$. To do this we must first determine the optimal feedback Hamiltonian. We note from Eq.(\ref{z1de}) that the measurement only introduces noise into the real part of $z_1$. So once the feedback Hamiltonian has been used to damp $\mbox{Im}[z_1]$ to zero it will stay zero, and so in the steady-state the optimal Hamiltonian is just that which damps the real part of $z_1$ as fast as possible. For a fixed maximum feedback strength $\mu = \mbox{Tr}[H^2]$, this just means choosing $H$ to be proportional to $\sigma_y$. Choosing linear feedback, which means that $H \propto z_1$, the feedback Hamiltonian is then 
\begin{equation} 
  H = \mbox{Re}[z_1] \sigma_y, 
\end{equation} 
and the resulting equation of motion is 
\begin{equation} 
   d \mbox{Re}[z_1]   =  -(k + \gamma + u ) \mbox{Re}[z_1] dt  + \sqrt{2k} dW . 
\end{equation} 
This stochastic differential equation is known as an Ornstein-Uhlenbeck equation, and solving it is straightforward using the standard techniques of Ito calculus~\cite{gardiner, JacobsSteck06}. The steady-state mean and variance are  
\begin{eqnarray}
    \langle \mbox{Re}[z_1]  \rangle_{\mbox{\scriptsize ss}} & = &  0,  \\   
    V_{\mbox{\scriptsize ss}} (\mbox{Re}[z_1] ) & = & \frac{k}{k + \gamma + u}  , 
    \label{vz} 
\end{eqnarray} 
where $V_{\mbox{\scriptsize ss}}$ denotes the variance. The conditions under which optimal linear feedback control preserves the regime of good control is $\lambda_1 \ll 1$ and $\sqrt{V_{\mbox{\scriptsize ss}} (\mbox{Re}[z_1] )} \ll 1$. Using Eqs.(\ref{lamss}) and (\ref{vz}) these conditions become 
\begin{eqnarray} 
   k & \gg &  \Gamma + \gamma, \label{con1} \\   
   \sqrt{u} & \gg  &  \sqrt{k} .  \label{con2} 
\end{eqnarray}
This is the complete solution to the steady-state feedback control problem for a single qubit undergoing decay and/or dephasing, with feedback from an unbiased measurement, in the regime of good control. 

\section{Discussion}

In this work we have defined the regime of good control as that in which, for the duration of the control, the relationship between the target state, $|\psi\rangle$, and the system density matrix, $\rho = \sum_{n=0}^{N-1} \lambda_n |n\rangle\langle n|$, is given by  
\begin{equation} 
       |\psi\rangle = (1-\Delta) |0\rangle +  \sum_{n=1}^{N-1} z_n |n\rangle ,  
\end{equation} 
where 
\begin{eqnarray}
   \lambda_n & \sim & \Delta \ll  1,     \;\;\;\; n \geq 1    \label{res1}  \\   
              z_n & \sim & \Delta  \ll 1,     \;\;\;\;  n \geq 1   \label{res2}  . 
\end{eqnarray}
That is, we have treated the eigenvalues of the density matrix, and the coefficients of the target state in the density matrix eigenbasis, $z_n$, on the same footing regarding the small parameter $\Delta$. This level of approximation preserves the linearity of the equations of motion for the coefficients $z_n$, and thus gives us the simplest description of feedback control. 

However, if we take a look at the conditions upon the measurement strength, $k$, and feedback strength, $u$, required to satisfy the regime of good control for a single qubit (Eqs.(\ref{con1}) and (\ref{con2})), we see that these conditions do not give measurement and feedback quite the same status. The condition given by Eq.(\ref{con1}), coming from the requirement on the eigenvalues, Eq.(\ref{res1}), involves a ratio of the rate constants. But the condition Eq.(\ref{con2}), coming from the requirement on the coefficients, Eq.(\ref{res2}), involves a ratio of the {\em square roots} of the rate constants. This means that the condition  $z_n \sim \Delta  \ll 1$, $n \geq 1$ is a {\em stronger} requirement than the eigenvalue condition $\lambda_n \sim \Delta  \ll 1$, $n \geq 1$. 
This comes from the fact that it is the {\em variance} of the coefficient $z_1$ that is proportional the ratio of the rate constants, and not the standard deviation of $z_1$. 

The above analysis implies that to impose the same level of constraint upon the eigenvalues and the coefficients, we must instead place the same requirement on the eigenvalues as we do on the {\em square moduli} of the coefficients. That is 
\begin{eqnarray}  
              |z_n|^2 & \sim & \Delta  \ll 1,     \;\;\;\;  n \geq 1 . 
\end{eqnarray} 
This less restrictive regime does not give linear equations for the coefficients $z_n$, but instead gives equations that depend also upon $z_n^2$. We will not investigate this regime further here, but we feel that it is an interesting subject for future work. 

Returning to the regime of good control as we have defined it here, we can summarize the results as follows.  The dynamics is described by two sets of variables, the  eigenvalues of the density matrix (giving $N-1$ independent real variables), and the elements of the eigenvectors (giving another $N^2$ independent real variables~\cite{Dita03}). How many of the elements of the eigenvectors are actually required depends on the noise, however. If the noise is completely independent of the eigenvectors (isotropic), or the system is a single qubit, then only the coefficients of the target state in the basis of the eigenvectors are required to describe the dynamics. In this case there are a total of only $2(N-1)$ real dynamical variables, where $N$ is the dimension of the system. 

The equations of motion for the eigenvectors (and thus the coefficients of the target state, and the elements of the noise operators) are linear, but the equations of motion for the eigenvalues are nonlinear for every system with dimension higher than two. This is in contrast to the regime of good control for classical systems. As the dimension of the system increases, the nonlinearity of the equations of motion for the eigenvalues increases, in that higher powers of the eigenvalues appear in these equations. This means that obtaining steady-state solutions to these equations involves solving increasingly high order polynomials as the system size increases. In addition to being nonlinear, the equations of motion also contain multiplicative noise for dimensions higher than two. 

In the example that we analyzed for a single qubit, we have considered only optimizing the feedback while fixing the measurement strategy.  Since the equations of motion in the regime of good control are linear for a single qubit, we expect that analytic expressions can be derived giving fully optimal feedback protocols for arbitrary noise, using results from classical control theory. Deriving and exploring these protocols is a natural topic for further work.


\begin{thebibliography}{32}
\expandafter\ifx\csname natexlab\endcsname\relax\def\natexlab#1{#1}\fi
\expandafter\ifx\csname bibnamefont\endcsname\relax
  \def\bibnamefont#1{#1}\fi
\expandafter\ifx\csname bibfnamefont\endcsname\relax
  \def\bibfnamefont#1{#1}\fi
\expandafter\ifx\csname citenamefont\endcsname\relax
  \def\citenamefont#1{#1}\fi
\expandafter\ifx\csname url\endcsname\relax
  \def\url#1{\texttt{#1}}\fi
\expandafter\ifx\csname urlprefix\endcsname\relax\def\urlprefix{URL }\fi
\providecommand{\bibinfo}[2]{#2}
\providecommand{\eprint}[2][]{\url{#2}}

\bibitem{OLRJacobs}
\bibinfo{author}{\bibfnamefont{O.~L.~R.} \bibnamefont{Jacobs}},
  \emph{\bibinfo{title}{Introduction to Control Theory}}
  (\bibinfo{publisher}{OUP, Oxford}, \bibinfo{year}{1993}).

\bibitem{Whittle96}
\bibinfo{author}{\bibfnamefont{P.}~\bibnamefont{Whittle}},
  \emph{\bibinfo{title}{Optimal Control}} (\bibinfo{publisher}{Wiley,
  Chichester}, \bibinfo{year}{1996}).

\bibitem{JacobsShabani09}
\bibinfo{author}{\bibfnamefont{K.}~\bibnamefont{Jacobs}} \bibnamefont{and}
  \bibinfo{author}{\bibfnamefont{A.}~\bibnamefont{Shabani}},
  \bibinfo{journal}{Contemp. Phys.}  (\bibinfo{year}{in press}).

\bibitem{JacobsSteck06}
\bibinfo{author}{\bibfnamefont{K.}~\bibnamefont{Jacobs}} \bibnamefont{and}
  \bibinfo{author}{\bibfnamefont{D.~A.} \bibnamefont{Steck}},
  \bibinfo{journal}{Contemp. Phys.} \textbf{\bibinfo{volume}{47}},
  \bibinfo{pages}{279} (\bibinfo{year}{2006}).

\bibitem{Brun02}
\bibinfo{author}{\bibfnamefont{T.~A.} \bibnamefont{Brun}},
  \bibinfo{journal}{{Am.\ J.\ Phys.}} \textbf{\bibinfo{volume}{70}},
  \bibinfo{pages}{719} (\bibinfo{year}{2002}).

\bibitem{Smith02}
\bibinfo{author}{\bibfnamefont{W.~P.} \bibnamefont{Smith}},
  \bibinfo{author}{\bibfnamefont{J.~E.} \bibnamefont{Reiner}},
  \bibinfo{author}{\bibfnamefont{L.~A.} \bibnamefont{Orozco}},
  \bibinfo{author}{\bibfnamefont{S.}~\bibnamefont{Kuhr}}, \bibnamefont{and}
  \bibinfo{author}{\bibfnamefont{H.~M.} \bibnamefont{Wiseman}},
  \bibinfo{journal}{{Phys.\ Rev.\ Lett.}} \textbf{\bibinfo{volume}{89}},
  \bibinfo{pages}{133601} (\bibinfo{year}{2002}).

\bibitem{Armen02}
\bibinfo{author}{\bibfnamefont{M.~A.} \bibnamefont{Armen}},
  \bibinfo{author}{\bibfnamefont{J.~K.} \bibnamefont{Au}},
  \bibinfo{author}{\bibfnamefont{J.~K.} \bibnamefont{Stockton}},
  \bibinfo{author}{\bibfnamefont{A.~C.} \bibnamefont{Doherty}},
  \bibnamefont{and} \bibinfo{author}{\bibfnamefont{H.}~\bibnamefont{Mabuchi}},
  \bibinfo{journal}{{Phys.\ Rev.\ Lett.}} \textbf{\bibinfo{volume}{89}},
  \bibinfo{pages}{133602} (\bibinfo{year}{2002}).

\bibitem{Steck04}
\bibinfo{author}{\bibfnamefont{D.}~\bibnamefont{Steck}},
  \bibinfo{author}{\bibfnamefont{K.}~\bibnamefont{Jacobs}},
  \bibinfo{author}{\bibfnamefont{H.}~\bibnamefont{Mabuchi}},
  \bibinfo{author}{\bibfnamefont{T.}~\bibnamefont{Bhattacharya}},
  \bibnamefont{and} \bibinfo{author}{\bibfnamefont{S.}~\bibnamefont{Habib}},
  \bibinfo{journal}{Phys. Rev. Lett.} \textbf{\bibinfo{volume}{92}},
  \bibinfo{pages}{223004} (\bibinfo{year}{2004}).

\bibitem{James04}
\bibinfo{author}{\bibfnamefont{M.~R.} \bibnamefont{James}},
  \bibinfo{journal}{Phys. Rev. A} \textbf{\bibinfo{volume}{69}},
  \bibinfo{pages}{032108} (\bibinfo{year}{2004}).
  
\bibitem{Ralph04} 
   J. F. Ralph, E. J. Griffith, T. D. Clark and M. J. Everitt, Phys. Rev. B {\bf 70}, 
   214521(2004). 

\bibitem{Combes06}
\bibinfo{author}{\bibfnamefont{J.}~\bibnamefont{Combes}} \bibnamefont{and}
  \bibinfo{author}{\bibfnamefont{K.}~\bibnamefont{Jacobs}},
  \bibinfo{journal}{Phys. Rev. Lett.} \textbf{\bibinfo{volume}{96}},
  \bibinfo{pages}{010504} (\bibinfo{year}{2006}).

\bibitem{Yanagisawa06}
\bibinfo{author}{\bibfnamefont{M.}~\bibnamefont{Yanagisawa}},
  \bibinfo{journal}{Phys. Rev. Lett.} \textbf{\bibinfo{volume}{97}},
  \bibinfo{pages}{190201} (\bibinfo{year}{2006}).

\bibitem{Geremia06}
\bibinfo{author}{\bibfnamefont{J.}~\bibnamefont{Geremia}},
  \bibinfo{journal}{Phys. Rev. Lett.} \textbf{\bibinfo{volume}{97}},
  \bibinfo{pages}{073601} (\bibinfo{year}{2006}).

\bibitem{Matsukevich06}
\bibinfo{author}{\bibfnamefont{D.~N.} \bibnamefont{Matsukevich}},
  \bibinfo{author}{\bibfnamefont{T.}~\bibnamefont{Chaneliere}},
  \bibinfo{author}{\bibfnamefont{S.~D.} \bibnamefont{Jenkins}},
  \bibinfo{author}{\bibfnamefont{S.-Y.} \bibnamefont{Lan}},
  \bibinfo{author}{\bibfnamefont{T.~A.~B.} \bibnamefont{Kennedy}},
  \bibnamefont{and} \bibinfo{author}{\bibfnamefont{A.}~\bibnamefont{Kuzmich}},
  \bibinfo{journal}{Phys. Rev. Lett.} \textbf{\bibinfo{volume}{97}},
  \bibinfo{pages}{013601} (\bibinfo{year}{2006}).

\bibitem{Bushev06}
\bibinfo{author}{\bibfnamefont{P.}~\bibnamefont{Bushev}},
  \bibinfo{author}{\bibfnamefont{D.}~\bibnamefont{Rotter}},
  \bibinfo{author}{\bibfnamefont{A.}~\bibnamefont{Wilson}},
  \bibinfo{author}{\bibfnamefont{F.}~\bibnamefont{Dubin}},
  \bibinfo{author}{\bibfnamefont{C.}~\bibnamefont{Becher}},
  \bibinfo{author}{\bibfnamefont{J.}~\bibnamefont{Eschner}},
  \bibinfo{author}{\bibfnamefont{R.}~\bibnamefont{Blatt}},
  \bibinfo{author}{\bibfnamefont{V.}~\bibnamefont{Steixner}},
  \bibinfo{author}{\bibfnamefont{P.}~\bibnamefont{Rabl}}, \bibnamefont{and}
  \bibinfo{author}{\bibfnamefont{P.}~\bibnamefont{Zoller}},
  \bibinfo{journal}{{Phys.\ Rev.\ Lett.}} \textbf{\bibinfo{volume}{96}},
  \bibinfo{pages}{043003} (\bibinfo{year}{2006}).

\bibitem{Ticozzi06}
\bibinfo{author}{\bibfnamefont{F.}~\bibnamefont{Ticozzi}} \bibnamefont{and}
  \bibinfo{author}{\bibfnamefont{L.}~\bibnamefont{Viola}},
  \bibinfo{journal}{Phys. Rev. A} \textbf{\bibinfo{volume}{74}},
  \bibinfo{pages}{052328} (\bibinfo{year}{2006}).

\bibitem{Tian07}
\bibinfo{author}{\bibfnamefont{L.}~\bibnamefont{Tian}}, \bibinfo{journal}{Phys.
  Rev. Lett.} \textbf{\bibinfo{volume}{98}}, \bibinfo{pages}{153602}
  (\bibinfo{year}{2007}).

\bibitem{Katz07}
\bibinfo{author}{\bibfnamefont{G.}~\bibnamefont{Katz}},
  \bibinfo{author}{\bibfnamefont{M.~A.} \bibnamefont{Ratner}},
  \bibnamefont{and} \bibinfo{author}{\bibfnamefont{R.}~\bibnamefont{Kosloff}},
  \bibinfo{journal}{Phys. Rev. Lett.} \textbf{\bibinfo{volume}{98}},
  \bibinfo{pages}{203006} (\bibinfo{year}{2007}).

\bibitem{Cook07}
\bibinfo{author}{\bibfnamefont{R.~L.} \bibnamefont{Cook}},
  \bibinfo{author}{\bibfnamefont{P.~J.} \bibnamefont{Martin}},
  \bibnamefont{and} \bibinfo{author}{\bibfnamefont{J.~M.}
  \bibnamefont{Geremia}}, \bibinfo{journal}{Nature}
  \textbf{\bibinfo{volume}{446}}, \bibinfo{pages}{774} (\bibinfo{year}{2007}).

\bibitem{Jacobs07c}
\bibinfo{author}{\bibfnamefont{K.}~\bibnamefont{Jacobs}} \bibnamefont{and}
  \bibinfo{author}{\bibfnamefont{A.~P.} \bibnamefont{Lund}},
  \bibinfo{journal}{Phys. Rev. Lett.} \textbf{\bibinfo{volume}{99}},
  \bibinfo{pages}{020501} (\bibinfo{year}{2007}).

\bibitem{Altafini07}
\bibinfo{author}{\bibfnamefont{C.}~\bibnamefont{Altafini}},
  \bibinfo{journal}{Quant. Inf. Proc.} \textbf{\bibinfo{volume}{6}},
  \bibinfo{pages}{9} (\bibinfo{year}{2007}).

\bibitem{Wang07}
\bibinfo{author}{\bibfnamefont{S.~K.} \bibnamefont{Wang}},
  \bibinfo{author}{\bibfnamefont{J.~S.} \bibnamefont{Jin}}, \bibnamefont{and}
  \bibinfo{author}{\bibfnamefont{X.~Q.} \bibnamefont{Li}},
  \bibinfo{journal}{Phys. Rev. B} \textbf{\bibinfo{volume}{75}},
  \bibinfo{pages}{155304} (\bibinfo{year}{2007}).

\bibitem{Sagawa08}
\bibinfo{author}{\bibfnamefont{T.}~\bibnamefont{Sagawa}} \bibnamefont{and}
  \bibinfo{author}{\bibfnamefont{M.}~\bibnamefont{Ueda}},
  \bibinfo{journal}{Phys. Rev. Lett.} \textbf{\bibinfo{volume}{100}},
  \bibinfo{pages}{080403} (\bibinfo{year}{2008}).

\bibitem{Combes08}
\bibinfo{author}{\bibfnamefont{J.}~\bibnamefont{Combes}},
  \bibinfo{author}{\bibfnamefont{H.~M.} \bibnamefont{Wiseman}},
  \bibnamefont{and} \bibinfo{author}{\bibfnamefont{K.}~\bibnamefont{Jacobs}},
  \bibinfo{journal}{Phys. Rev. Lett.} \textbf{\bibinfo{volume}{100}},
  \bibinfo{pages}{160503} (\bibinfo{year}{2008}).

\bibitem{Maybeck3}
\bibinfo{author}{\bibfnamefont{P.~S.} \bibnamefont{Maybeck}},
  \emph{\bibinfo{title}{Stochastic Models, Estimation and Control}}, vol.
  \bibinfo{volume}{III} (\bibinfo{publisher}{Academic Press, New York},
  \bibinfo{year}{1982}).

\bibitem{Shabani08}
\bibinfo{author}{\bibfnamefont{A.}~\bibnamefont{Shabani}} \bibnamefont{and}
  \bibinfo{author}{\bibfnamefont{K.}~\bibnamefont{Jacobs}},
  \bibinfo{journal}{Phys. Rev. Lett.} {\bf 101}, 230403 (2008).

\bibitem{rapidP}
\bibinfo{author}{\bibfnamefont{K.}~\bibnamefont{Jacobs}},
  \bibinfo{journal}{Phys. Rev. A} \textbf{\bibinfo{volume}{67}},
  \bibinfo{pages}{030301(R)} (\bibinfo{year}{2003}).

\bibitem{Wiseman06x}
\bibinfo{author}{\bibfnamefont{H.~M.} \bibnamefont{Wiseman}} \bibnamefont{and}
  \bibinfo{author}{\bibfnamefont{J.~F.} \bibnamefont{Ralph}},
  \bibinfo{journal}{New. J. Phys} \textbf{\bibinfo{volume}{8}},
  \bibinfo{pages}{90} (\bibinfo{year}{2006}).

\bibitem{Wiseman07}
\bibinfo{author}{\bibfnamefont{H.~M.} \bibnamefont{Wiseman}} \bibnamefont{and}
  \bibinfo{author}{\bibfnamefont{L.}~\bibnamefont{Bouten}},
  \bibinfo{journal}{Quant. Inf. Proc.} \textbf{\bibinfo{volume}{7}},
  \bibinfo{pages}{71} (\bibinfo{year}{2007}).

\bibitem{Bransden}
\bibinfo{author}{\bibfnamefont{B.~H.} \bibnamefont{Bransden}} \bibnamefont{and}
  \bibinfo{author}{\bibfnamefont{C.~J.} \bibnamefont{Joachain}},
  \emph{\bibinfo{title}{Introduction to Quantum Mechanics}}
  (\bibinfo{publisher}{Longman, Essex}, \bibinfo{year}{1989}).

\bibitem{DHJMT}
\bibinfo{author}{\bibfnamefont{A.~C.} \bibnamefont{Doherty}},
  \bibinfo{author}{\bibfnamefont{S.}~\bibnamefont{Habib}},
  \bibinfo{author}{\bibfnamefont{K.}~\bibnamefont{Jacobs}},
  \bibinfo{author}{\bibfnamefont{H.}~\bibnamefont{Mabuchi}}, \bibnamefont{and}
  \bibinfo{author}{\bibfnamefont{S.~M.} \bibnamefont{Tan}},
  \bibinfo{journal}{Phys. Rev. A} \textbf{\bibinfo{volume}{62}},
  \bibinfo{pages}{012105} (\bibinfo{year}{2000}).

\bibitem{Wiseman01}
\bibinfo{author}{\bibfnamefont{H.~M.} \bibnamefont{Wiseman}} \bibnamefont{and}
  \bibinfo{author}{\bibfnamefont{L.}~\bibnamefont{Diosi}},
  \bibinfo{journal}{Chem. Physics} \textbf{\bibinfo{volume}{91}},
  \bibinfo{pages}{268} (\bibinfo{year}{2001}).
  
\bibitem{ZhouDoyle97}
  K. Zhou and J. C. Doyle, {\em Essentials of Robust Control} (Prentice Hall, 1997). 

\bibitem{gardiner}
\bibinfo{author}{\bibfnamefont{C.~W.} \bibnamefont{Gardiner}},
  \emph{\bibinfo{title}{Handbook of Stochastic Methods}}
  (\bibinfo{publisher}{Springer}, \bibinfo{year}{1985}).
  
\bibitem{Dita03} 
    P. Di{\c{t}}{\u{a}}, J. Phys. A {\bf 36}, 2781 (2003). 

\end{thebibliography}
\end{document}